\documentclass[11pt]{article}
\usepackage{amssymb}
\usepackage{amsmath}
\usepackage{amsfonts}
\makeatletter
\def\eqnarray{\stepcounter{equation}\let\@currentlabel=\theequation
\global\@eqnswtrue
\global\@eqcnt\z@\tabskip\@centering\let\\=\@eqncr
$$\halign to \displaywidth\bgroup\@eqnsel\hskip\@centering
  $\displaystyle\tabskip\z@{##}$&\global\@eqcnt\@ne
  \hfil$\displaystyle{\hbox{}##\hbox{}}$\hfil
  &\global\@eqcnt\tw@ $\displaystyle\tabskip\z@
  {##}$\hfil\tabskip\@centering&\llap{##}\tabskip\z@\cr}
\@addtoreset{equation}{section}
  \def\theequation{\thesection.\arabic{equation}}
\makeatother
\begin{document}

\title{Bosonized supersymmetry from the
Majorana-Dirac-Staunton theory and massive higher-spin fields}

\author{
{\sf Peter A. Horv\'athy${}^a$}\footnote{ E-mails:
horvathy@lmpt.univ-tours.fr}, {\sf Mikhail S.
Plyushchay${}^b$}\footnote{mplyushc@lauca.usach.cl},
{\sf Mauricio Valenzuela${}^b$}\footnote{mauricio.valenzuela@correo.usach.cl}\\
[4pt] {\small \it ${}^a$Laboratoire de Math\'ematiques et de
Physique Th\'eorique, Universit\'e de Tours,}\\
{\small \it Parc de Grandmont,
 F-37200 Tours, France}\\
{\small \it ${}^b$Departamento de F\'{\i}sica, Universidad de
Santiago de Chile, Casilla 307, Santiago 2, Chile} }

\date{}

\maketitle

\begin{abstract}
We propose a $(3+1)D$ linear set of covariant vector equations,
which unify the spin $0$ ``new Dirac equation'' with its  spin
$1/2$ counterpart, proposed by Staunton. Our equations describe a
spin $(0,1/2)$ supermultiplet with different numbers of degrees of
freedom in the bosonic and fermionic sectors. The
translation-invariant spin deegres of freedom are carried by two
copies of the Heisenberg algebra. This allows us to realize
space-time supersymmetry in a bosonized form. The grading
structure is provided by an internal reflection operator. Then the
construction is generalized by means of the Majorana equation to a
supersymmetric theory of massive higher-spin particles. The
resulting theory is characterized by a nonlinear symmetry
superalgebra, that, in the large-spin limit, reduces to the
super-Poincar\'e algebra with or without tensorial central charge.
\end{abstract}

\section{Introduction}

In 1932, Ettore Majorana \cite{Maj,Frad} proposed a Lorentz
invariant linear differential equation, associated with infinite
dimensional unitary representations of the Lorentz group. The
subsequent development of the concept of the infinite component
fields \cite{Fro1}--\cite{Stoyanov-Todorov} culminated in the
construction of the dual resonance models, and lead eventually to
superstring theory  \cite{Venez}--\cite{Green:1980zg}.

The Majorana equation has massive, massless and tachyonic
solutions (see Refs. \cite{Plyushchay:2006pw,Casalbuoni:2006fa}
for recent reviews). In the massive case, the equation describes
two series of positive-energy particles with arbitrary integer or
half-integer spin. The equation doesn't fix the mass, however,
rather provides a spin-dependent, Regge-like mass spectrum,
(\ref{MassSpectr}) below\footnote{ For a discussion of the
spin-statistics relation for the Majorana field, see  Ref.
\cite{Stoyanov-Todorov,Sudarshan:1970ss}.}. The simultaneous
presence of integer and half-integer spins suggests, together with
the positivity of the energy, that some kind of
\emph{supersymmetry} could be involved in the Majorana
construction \cite{Plyushchay:2006pw,Zichichi}.

In 1971, Dirac \cite{Dirac} put forward a linear spinor set of
equations, from which the Majorana and Klein-Gordon equations
follow as integrability (consistency) conditions. This ``new Dirac
equation'' 
describes massive particles with zero spin.

 A couple of years later, Staunton \cite{Stau2}
proposed, instead of the spinorial Dirac approach, a  vector
equation, which involves a new parameter, $\kappa$. Staunton's new
system is only consistent for $\kappa=1/2$ or $1$. For $\kappa=1/2$,
his equation coincides with one of the consistency relations implied
by the  equations of Dirac; it describes hence a spin $0$ massive
particle. The second value, $\kappa=1$, yields a spin $1/2$ particle
of nonzero mass. The Staunton equations imply, once again, the
Klein-Gordon and Majorana equations as integrability conditions.
With some abuse of language, the $\kappa=1/2$ (i.e. spin $0$)
equation of Staunton will be referred to as ``the new Dirac system''
(to which it is equivalent), and the $\kappa=1$ (i.e. spin $1/2$)
equation  will be referred to as ``the Staunton system''.

The Dirac and Staunton solutions both have positive
energy. Their masses can also  be derived from the Majorana
spectrum  [(\ref{MassSpectr}) below] with appropriate mass parameters, out
of which those solutions which carry the lowest possible spin,
namely spin-$0$ and spin-$1/2$, respectively, were selected.

In this paper we show first that the Dirac and Staunton equations
can be merged into a single supersymmetric system. Then, with the
help of the modified Majorana equation, we generalize the
construction to a supersymmetric theory of massive higher-spin
particles. It is worth stressing that supersymmetrization is
achieved here without enlarging the system by adding new degrees
of freedom, as it is done usually. The necessary degrees of
freedom have already been present in the Dirac-Staunton and
Majorana frameworks. The underlying space is in fact decomposed
into two subspaces and the Dirac and resp. Staunton equations
merely select one sector and kill the other. Our unified system
simply activates them simultaneously.

 The two subspaces of the
Majorana equation  are initially
 unrelated.  A smart
choice of the mass parameter, however, creates a supersymmetry
between the two sectors.  This is similar to what happens for a
planar anisotropic oscillator, for which rational tuning of the
frequency ratio generates a (nonlinear) symmetry \cite{Boer}.

Since we only use bosonic variables, what we get here
is \emph{bosonized supersymmetry}. The unification of the even
and odd spin representations  and their supersymmetry relies
on using a non-local operator, namely the \emph{
reflection operator}.

Examples in which supersymmetry is realized  within a purely
bosonic system were presented recently in $D<4$ dimensions
\cite{Plyushchay:1994re,bosSUSY,anyonsusy}. For all these systems,
the r\^ole of the grading operator is played by the non-local
reflection operator.

 The present paper extends these results to
$(3+1)$ dimensions.

In the theories of Majorana, Dirac, and Staunton the field
equations involve, in their internal structure, two copies of the
Heisenberg algebra, associated with an internal planar harmonic
oscillator, $[q_i,\eta_i]=i\delta_{ij}, i=1,2$, as well as ten
quadratic products built out of these generators. Six quadratic
combinations span the Lorentz algebra.  The remaining four form a
Lorentz vector. These ten generators span, together, the anti
de-Sitter $so(2,3)$ algebra (analogously as Dirac matrices and
their commutators do). The Heisenberg algebra generators
$q_i,\,\eta_i, \, i=1,2$, can be united into a four-component
operator, say $L_a$. The latter transforms covariantly (namely as
a spinor) under the action of
 $so(2,3)$  and provides
us with a (bosonized) representation of the superalgebra $osp(1|4)$.

Then we can build the reflection operator ${\cal R}=(-1)^{(N_1 +
N_2)},$ where $N_1$ and $N_2$ are the number operators of the
Heisenberg algebras.  ${\cal R}$ commutes with the $so(2,3)$
subalgebra, anticommutes with the supercharge $L_a$ and has
eigenvalues $\pm1$. It provides us therefore with the grading
operator of $osp(1|4)$. The operator ${\cal R}$ can be identified
with a certain class of finite $SO(2,3)$ transformations,
 namely with internal reflection $q_i\mapsto -q_i$, or
alternatively, a nonlocal, finite  rotation (by $\pi$) in the 2D
plane, spanned by the $q_i$.

Technically, the unification of the Dirac and Staunton equations
boils down to first promoting Staunton's parameter $\kappa$ into an
operator by  inserting the reflection operator, ${\cal R}$, Eqn.
(\ref{kappa}) below, and then putting $\hat{\kappa}$ into Staunton's
general equation. On the $\pm1$ eigenspaces of  ${\cal R}$,
$\hat{\kappa}$ takes precisely the correct ``Dirac'' and
``Staunton'' values, $\kappa=1/2$ and $1$, respectively. The
restriction of our new equation reproduces, therefore, the spinless
Dirac and the spin $1/2$ Staunton equations, projected into the
corresponding eigenspaces of ${\cal R}$.

The two (namely spin-$0$ and spin-$\frac{1}{2}$) sectors can be
related by a Hermitian supercharge operator $Q_a$, which carries a
spin $\frac{1}{2}$ representation of the Lorentz group. As a
result, we get a nonlinear extension of the usual super-Poincar\'e
algebra by non-abelian tensor conserved charges, which appear in
the anticommutator of the supercharge.

Then we construct a generalized Majorana equation that provides us
with a supersymmetric system of fields  with spins $(j,j+1/2)$. In
the generic case of integer or half-integer $j$, such a system is
described by a $(3+1)D$ bosonized supersymmetry whose form has
been slightly modified when compared to the simplest, $j=0$, case.

The generalization is achieved in  a way
similar to the one we followed for the Dirac-Staunton theory:
we modify the mass parameter
in the original Majorana equation by introducing into it
 the operator ${\cal R}$ in a way that guarantees  that
the spin $j$ and $j+1/2$ states have equal masses. Then requiring
that a supercharge should exist and act as a symmetry
implies the Klein-Gordon equation as consistency condition.
 As a result, we obtain  a bosonized supersymmetric
theory of massive higher-spin particles, characterized by a
nonlinear superalgebraic structure. In the large-spin limit the
nonlinearity disappears, and the usual super-Poincar\'e algebra
with or without tensorial central charge
\cite{Bergshoeff:1987cm}-\cite{Bandos:1999rp} is recovered.

The paper is organized as follows. In Section \ref{intdeg}, we
construct, starting with two Heisenberg algebras, an
infinite-dimensional, unitary representation of the $osp(1|4)$
superalgebra and the reflection operator ${\cal R}$. In Section
\ref{MajDirStau}, we give a brief review of the theories of
Majorana, of Dirac and of Staunton. The supersymmetric theory for
the spin $(0,\frac{1}{2})$ supermultiplet is developed in Section
\ref{ourtheory}, where  the supersymmetric field
equation and the corresponding superalgebra are constructed.

 These results are
extended to an arbitrary spin supermultiplet by means of a
generalized Majorana equation in section \ref{Majoranasusy}.

Section \ref{conc}
includes comments and concluding remarks.

\section{Majorana representation and $osp(1|4)$}\label{intdeg}
The Majorana representation of the Lorentz group is an infinite
dimensional representation in which the Casimir operators,
\begin{equation}\label{C1C2}
C_1=S^{\mu\nu}S_{\mu\nu} \quad\hbox{and}\quad
C_2=\epsilon^{\mu\nu\lambda\rho}S_{\mu\nu}S_{\lambda\rho}\,,
\end{equation}
take the fixed  values
\begin{equation}\label{C1C2V}
C_1=-\frac{3}{2},\qquad C_2=0.
\end{equation}
This representation can be realized in terms of two copies of
Heisenberg algebras.

The Majorana representation
can be embedded into a larger  supersymmetric structure,
namely into $osp(1|4)$.
Let us indeed consider the two dimensional Heisenberg algebra generated by
the operators $q_i$ and $\eta_j$,
$$
\big[q_i,\eta_j\big]=i\delta_{ij}.
$$
We assume the coordinates $q_i$ are rescaled by a length parameter
$l$ so that the  generators  $q_i$ and $\eta_i$ are dimensionless.
The four-component operator
\begin{equation}
(L_a)=(q_1,q_2,\eta_1,\eta_2), \qquad a=1,2,3,4,
 \label{La}
\end{equation}
satisfies the relation
\begin{eqnarray}
[L_a,L_b]=iC_{ab},
\qquad
C_{ab}=
\left(%
\begin{array}{cc}
  0 & I_{2\times2} \\
  -I_{2\times2} & 0 \\
\end{array}
\label{Cab} \right).
\end{eqnarray}
The antisymmetric matrix $C_{ab}$ here can be viewed as a
metric tensor in the spinor indices, see below. Defining
$C^{ab}=C_{ab}$, $C_{ac}C^{bc}=\delta_a^b$, we rise and lower
indices as $L^a=L_bC^{ba}$ and $L_a=C_{ab}L^b$.

Ten independent tensor products $L_aL_b$ can be constructed and
combined as
\begin{equation}\label{so32g}
S_{\mu\nu}=\frac{i}{2}L^a(\gamma_{\mu \nu})_a\,^b L_b,\qquad
\Gamma_\mu=\frac{1}{4}L^a(\gamma_\mu)_a\,^b L_b,\qquad
\mu=0,1,2,3.
\end{equation}

Here the
$(\gamma ^\mu)_a\,^b$ are the Dirac matrices in the Majorana
representation,
$$
{\small
(\gamma^0)_a\,^b=\left(%
\begin{array}{cc}
  0 & \sigma^0 \\
  -\sigma^0 & 0 \\
\end{array}%
\right),\qquad
(\gamma^1)_a\,^b=\left(%
\begin{array}{cc}
  0 & \sigma^0 \\
  \sigma^0 & 0 \\
\end{array}%
\right)},$$
$$
{\small
(\gamma^2)_a\,^b=\left(%
\begin{array}{cc}
  \sigma^3 & 0 \\
  0 & -\sigma^3 \\
\end{array}\right),\qquad
(\gamma^3)_a\,^b=\left(%
\begin{array}{cc}
  -\sigma^1 & 0 \\
  0 & \sigma^1 \\
\end{array}%
\right)},$$ and $\gamma^{\mu\nu}=-\frac{i}{4}[\gamma^\mu,\gamma
^\nu]$ are pure imaginary matrices. Dirac matrices satisfy, with
the space-time metric $diag(\eta^{\mu\nu})=(-+++)$, the relation
$\gamma^\mu \gamma^\nu=\eta^{\mu\nu}+2i\gamma^{\mu\nu}$.

The quadratic operators (\ref{so32g}) and the $L_a$ generate the
$osp(1|4)$ superalgebra,
\begin{eqnarray}
&[S_{\mu\nu},S_{\lambda\rho}]=i(\eta_{\mu\lambda}S_{\nu\rho}+\eta_{\nu\rho}S_{\mu\lambda}-
\eta_{\mu\rho}S_{\nu\lambda}-\eta_{\nu\lambda}S_{\mu\rho}),&\label{Lorentz}
 \\[8pt]
&[S_{\mu\nu},\Gamma_\lambda]=i(\eta_{\mu\lambda}\Gamma_\nu-\eta_{\nu\lambda}\Gamma_\mu),
\qquad[\Gamma_\mu,\Gamma_\nu]=-iS_{\mu\nu},&\label{SG}
\\[8pt]
&[S_{\mu\nu},L_a]=-(\gamma_{\mu\nu})_a\,^bL_b,\qquad[\Gamma_\mu,L_a]=\frac{i}{2}(\gamma_\mu)_a\,^bL_b,&
\\[8pt]
&\{L_a,L_b\}=-2(iS_{\mu\nu}\gamma^{\mu\nu}-\Gamma_\mu\gamma^{\mu})_{ab},&\label{LL}
\end{eqnarray}
where $(\gamma^{\mu})_{ab}=C_{bc}(\gamma^\mu)_a\,^c$ and
$(\gamma^{\mu\nu})_{ab}=C_{bc}(\gamma^{\mu\nu})_a\,^c$ are symmetric
matrices.

The usual creation and annihilation operators are obtained from the
linear combinations $a^\pm _i=\frac{1}{\sqrt{2}}(q_i \mp
i\eta_i)$,\,\,$[a_i^-,a_j^+]=\delta_{ij},\,\, i,j=1,2$. So, the
$osp(1|4)$ generators act irreducibly on the tensor product of the
two Fock spaces,
\begin{equation}\label{Fock}
{\cal O} = \{ \vert n_1,\,n_2\rangle = \vert n_1\rangle  \vert
n_2\rangle ,\, n_1,n_2=0,1,2,...\},
\end{equation}
upon which the annihilation and creation operators act as
\begin{eqnarray}
    a^+_1\,\vert n_1, n_2\rangle =\sqrt{n_1+1}\,\vert n_1+1, n_2\rangle,\quad
    a^+_2\,\vert n_1, n_2\rangle =\sqrt{n_2+1}\,\vert n_1, n_2+1\rangle,
    \\[10pt]
    a^-_1\,\vert n_1, n_2\rangle =\sqrt{n_1}\,\vert n_1-1, n_2\rangle,\quad
    a^-_2\,\vert n_1, n_2\rangle =\sqrt{n_2}\,\vert n_1, n_2-1\rangle.
\end{eqnarray}
Here
\begin{equation}\label{N}
    N_1\,\vert n_1, n_2\rangle=n_1\,\vert n_1, n_2\rangle,\qquad
    N_2\,\vert n_1, n_2\rangle=n_2\,\vert n_1, n_2\rangle
\end{equation}
are the number operators $N_1=a^+_1a^-_1$ and $N_2=a^+_2a^-_2$,
respectively.

The $so(2,3)$ subalgebra (\ref{Lorentz})-(\ref{SG}) acts,
instead, reducibly over the whole space ${\cal O}$. Its
irreducible representations are spanned by the subspaces
\begin{equation}
{\cal O}_+={\vert ++\rangle \bigoplus \vert --\rangle}
\qquad\hbox{and}\qquad {\cal O}_-= {\vert+-\rangle\bigoplus \vert
-+\rangle}, \label{O+O-}
\end{equation}
where we defined
\begin{eqnarray}
&\vert{\pm}{\pm}\rangle = \{ \,\vert n_1,n_2\rangle _{{\pm}{\pm}}
= \vert n_1\rangle_{\pm}\,\vert n_2\rangle_{\pm},\,
n_1,n_2=0,1,2,...\},&\label{even}
\\[8pt]
&\vert {\pm}{\mp}\rangle = \{ \,\vert n_1,n_2\rangle _{{\pm}{\mp}}
= \vert n_1\rangle_{\pm}\,\vert n_2\rangle_{\mp},\,
n_1,n_2=0,1,2,...\}&\label{odd},
\\[8pt]
&
\qquad \vert n\rangle_+=\vert 2n\rangle,\qquad \vert
n\rangle_-=\vert 2n+1\rangle.&\label{n+-}
\end{eqnarray}
In this representation the Casimir operators ({\ref{C1C2}) take
the same values (\ref{C1C2V}) in both subspaces ${\cal O}_+$ and
${\cal O}_-$. Moreover, the square of the vector operator
$\Gamma_\mu$ is Lorentz invariant and is also fixed here,
\begin{equation}\label{G12}
    \Gamma^\mu\Gamma_\mu=\frac {1}{2}.
\end{equation}
We also have the identities
\begin{eqnarray}\label{id}
\Gamma^\mu
S_{\mu\nu}=S_{\nu\mu}\Gamma^\mu=-\frac{3i}{2}\Gamma_\nu,\qquad
\epsilon^{\mu\nu\lambda\rho}S_{\nu\lambda}\Gamma_\rho=0.
\end{eqnarray}

The operators $R_i=(-1)^{N_i}=\cos(\pi N_i),\, i=1,2$ are defined
in terms of the number operators (\ref{N}).
Acting on ${\cal O}$, they produce
$$
R_1 \vert n_1, n_2 \rangle=(-1)^{n_1} \vert n_1, n_2\rangle,
\qquad R_2 \vert n_1, n_2\rangle=(-1)^{n_2} \vert n_1, n_2\rangle.
$$
Then we introduce the \emph{total reflection
operator}
\begin{equation}\label{R}
    {\cal R}=R_1 R_2=(-1)^{N_1 + N_2}.
\end{equation}
In accordance with (\ref{O+O-}) and (\ref{R}),
\begin{equation}\label{RO}
    {\cal RO}_\pm=\pm{\cal O}_\pm, \qquad {\cal R}^2=1.
\end{equation}
${\cal R}$ plays the r\^ole of the grading operator in the
$osp(1|4)$ superalgebra (\ref{Lorentz})-(\ref{LL}): the relation
$\{{\cal R},a^\pm_i\}=0$ implies
\begin{equation}\label{grad}
[{\cal R},S_{\mu\nu}]=0, \qquad [{\cal R},\Gamma_\mu]=0,\qquad
\{{\cal R},L_a\}=0.
\end{equation}

We notice that, on account of the identity $(-1)^{2N_2}=1$ and the
explicit form of the AdS generators (see Appendix), the reflection
operator can be identified with two specific finite
transformations,
\begin{equation}\label{RR'}
{\cal R}=-\exp(i2\pi \Gamma_0)=\exp(i2\pi S^{12}),
\end{equation}
i.e., an AdS $2\pi$ rotation in the subspace of two time-like
coordinates, and a $2\pi$ space rotation, respectively. Since any
unitary $SO(2,3)$ transformation $U$ commutes with ${\cal R}$, we
have, more generally,
 ${\cal R}=-\exp(i2\pi \tilde{\Gamma}_0)=\exp(i2\pi
\tilde{S}^{12})$, where $\tilde{\Gamma}_0=U\Gamma_0U^\dagger$,
$\tilde{S}^{12}=U{S}^{12}U^\dagger$.  In any case, the reflection
operator, being a $\pi$-rotation in the 2D plane spanned by the
coordinates $q_i$, is non-local.  In the corresponding Schr\"odinger
representation
\begin{equation}\label{reflection}
{\cal R}\psi(\vec{q}\,)=\psi(-\vec{q}\,).
\end{equation}
The eigenfunctions of ${\cal R}$ are therefore either even or odd,
\begin{equation}
{\cal R}\psi_\pm(\vec{q}\,)=\pm \psi_\pm(\vec{q}\,),\qquad
\psi_\pm(\vec{q}\,)=\frac{1}{2}(
\psi(\vec{q}\,)\pm\psi(-\vec{q}\,)).
\end{equation}

\section{The relativistic wave equations of
Majorana, Dirac and of Staunton}
\label{MajDirStau}

In this Section we briefly review the Majorana  equation
\cite{Maj}, together with the related systems of spinor and vector
  equations proposed by Dirac \cite{Dirac}, and
by Staunton \cite{Stau2}.

\subsection{The Majorana equation}
The  Majorana  equation   \cite{Maj} is a Lorentz invariant
equation based on the unitary infinite dimensional (reducible)
representation of the  AdS algebra  described above,
\begin{equation}\label{maj}
(P^\mu\Gamma_\mu-M)|\Psi(x)\rangle=0.
\end{equation}
Here the $x^\mu$ are the space-time coordinates,
$P_\mu=-i\partial/\partial x^\mu$, and
$S_{\mu\nu}=i[\Gamma_\mu,\Gamma_\nu]$ is the translation-invariant
part of the Lorentz generators
\begin{equation}
{\cal J}_{\mu\nu}=x_{\mu}P_\nu-x_{\nu}P_\mu+S_{\mu\nu}.
\end{equation}
Since the AdS algebra acts irreducibly only in the subspaces ${\cal
O}_+$ and ${\cal O}_-$ of the internal Fock space,
$|\Psi(x)\rangle=|\Psi_+(x) \rangle+|\Psi_-(x) \rangle$ is an
infinite-component field expanded in these subspaces,
\begin{equation}\label{maj:sol}
|\Psi_\pm(x) \rangle=\sum_{\,\,{\cal
O}_\pm}\psi^\pm_{n_1,n_2}(x)|n_1,n_2\rangle,
\end{equation}
where  the ``$\pm$" label indicates that the field has been
expanded over the $\pm$ eigenspaces of ${\cal R}$,
\begin{equation}\label{R:psi}
 {\cal R}|\Psi_\pm(x) \rangle=\pm
|\Psi_\pm(x) \rangle, \qquad |\Psi_\pm(x)
\rangle\equiv\Pi_\pm|\Psi(x)\rangle .
\end{equation}
Here we have introduced the projectors
\begin{eqnarray}\label{projectors}
\Pi_+=\frac{1}{2}(1+{\cal R}), \qquad
\Pi_-=\frac{1}{2}(1-{\cal R}),\\[12pt]
\Pi_++\Pi_-=1,\qquad\left(\Pi_\pm\right)^2=\Pi_\pm,\qquad
\Pi_+\Pi_-=0.\nonumber
\end{eqnarray}
Note that
\begin{equation}\label{Pi,SGL}
[\Pi_\pm,S_{\mu \nu}]=[\Pi_\pm,\Gamma_{\mu}]=0,\qquad \Pi_\pm
L_a=L_a \Pi_\mp.
\end{equation}}

The square of the spin vector, built out of the space part of
$S_{\mu\nu}$, $S_i=\frac{1}{2}\epsilon_{ijk}S_{jk}$, is
\begin{eqnarray}
S_iS_i=\hat{J}(\hat{J}+1),
\end{eqnarray}
where
\begin{eqnarray}
\hat{J}=\frac{N_1+N_2}{2}\, .
\label{JNdef}
\end{eqnarray}
$\hat{J}$  is related to the AdS operator $\Gamma_0$ by
\begin{eqnarray}
\Gamma_0=\hat{J}+\frac{1}{2}.
 \label{gamma0J}
\end{eqnarray}

The Majorana equation (\ref{maj}) has  massive, massless and
tachyonic solutions. Below we restrict our analysis to the massive
sector. Passing to the rest frame, we put $P^\mu=(m_J,0,0,0)$ in
(\ref{maj}).  Then using (\ref{gamma0J}), we obtain the  celebrated
$J$-dependent mass spectrum,
\begin{equation}
m_J=\frac{M}{(J+\frac{1}{2})}, \label{MassSpectr}
\end{equation}
where $J$, the spin, is the eigenvalue of $\hat{J}$ acting over the
physical subspace.

The Majorana equation admits two independent sets of solutions,
composed of integer and of half-integer spins, respectively. These
values correspond precisely to the eigen-subspaces ${\cal O}_+$
and ${\cal O}_-$ of the reflection operator, see (\ref{RO}). This
follows from
\begin{equation}\label{R:J}
{\cal R}=(-1)^{2\hat{J}},
\end{equation}
inferred from (\ref{JNdef}) and (\ref{R}).

The solutions (\ref{maj:sol}) of the Majorana equation
are superpositions of those solutions which carry spin $J_\pm$
and mass
$m_{J_\pm}$,
\begin{eqnarray}
|\Psi_\pm(x) \rangle=\sum_{J_\pm}|\Psi_{J_\pm}(x)\rangle,
\qquad 
J_+=0,1,2,...\,,\quad
J_-={^1}/{_2},{^3}/{_2},{^5}/{_2},...\,\,.\nonumber
\end{eqnarray}
Consistently, the square of the Pauli-Lubanski vector
$W^\mu=\frac{1}{2}\epsilon^{\mu\nu\lambda\rho}S_{\nu\lambda}P_\rho$,
\begin{eqnarray}\label{Wsq1}
    W^\mu W_\mu=-\frac{1}{2}S^{\mu\nu}S_{\mu\nu}P^2+S_{\mu\nu}S^{\mu\lambda}P^\nu
    P_\lambda
    =\frac{1}{4}P^2+(P\Gamma)^2,
\end{eqnarray}
takes, when restricted to  these states, the (on-shell) value
\begin{equation}\label{Wsq2}
    W^\mu W_\mu|\Psi_{J}(x)\rangle=m_J^2 J(J+1)|\Psi_{J}(x)\rangle.
\end{equation}
In this way, in the massive sector $P^2<0$ the Majorana equation
(\ref{maj}) describes an infinite sum of irreducible representations
of the Poincar\'e group of arbitrary spin $J$ and of mass $m_J$
related via Eq. (\ref{MassSpectr}). At the same time we can expand
\begin{eqnarray}
    |\Psi_{J_+}(x)\rangle= &\sum_{n_1,n_2=0}^\infty \left(\psi^{n_1 n_2}_{++}(x)\vert n_1,n_2
    \rangle_{++}+\psi^{n_1 n_2}_{--}(x)\vert n_1 ,n_2 \rangle_{--}\right),\qquad
    &\hbox{ in}\quad \cal{O}_+ \, ,\label{psi+}
    \\[6pt]
    |\Psi_{J_-} (x)\rangle=&\sum_{n_1,n_2=0}^\infty \left(\psi^{n_1 n_2}_{+-}(x)\vert n_1 ,n_2
    \rangle_{+-}+\psi^{n_1 n_2}_{-+} (x)\vert n_1
    ,n_2\rangle_{-+}\right),
    &\hbox{ in}\quad \cal{O}_- .\label{psi-}
\end{eqnarray}
In the rest-frame,
\begin{equation}\label{O+}
|\Psi_{J_+}^{(0)}(x)\rangle=\left(\sum_{n=0}^{J_+}\psi^n_{++}\vert
J_+-n,n\rangle_{++} +\sum_{n=1}^{J_+}\psi^n_{--}\vert
J_+-n,n-1\rangle_{--}\right)\exp(-itm_{J_+}),
\end{equation}
\begin{equation}\label{O-}
|\Psi_{J_-}^{(0)}(x)\rangle=\sum_{n=0}^{J_--\frac{1}{2}}\left(\psi^n_{+-}\vert
J_--\frac{1}{2}-n,n\rangle_{+-}+\psi^n_{-+}\vert
J_--\frac{1}{2}-n,n\rangle_{-+}\right)\exp(-itm_{J_-}),
\end{equation}
where the $\psi^n_{\pm\pm}$, $\psi^n_{\pm\mp}$ are arbitrary
constants and $t=x^0$. These expansions correspond to the
superposition of the $2J+1$ possible polarization states (with
$J=J_+$ or $J=J_-$). Every state is an eigenvector of the operator
$S_z=S_{12}=\frac{1}{2}(N_1-N_2)$, which is the projection of the
spin on the z-axis; it has eigenvalues $\{-J,-J+1,...,J-1 ,J\}$.
(\ref{psi+}) and (\ref{psi-})  can be obtained  by a suitable
 Lorentz transformation of
(\ref{O+}) and (\ref{O-}). Hence, only $2J+1$ components are
independent in (\ref{psi+}) and (\ref{psi-}).

Because both ${\cal O}_+$ and ${\cal O}_-$ carry irreducible
representations of the
 Lorentz group, the solutions of
integer and half-integer spin are, in principle, independent.
We
can make an important observation, however. The direct sum ${\cal
O}_+\bigoplus{\cal O}_-$ spans an irreducible representation of the
$osp(1|4)$ superalgebra, where the spinor supercharge operator,
$L_a$, interchanges the subspaces~: $L_a:{\cal
O}_+\longleftrightarrow{\cal O}_-$.

So we have the possibility to construct a (super)symmetry based on
the $L_a$ operators, which connect the solutions of the Majorana
equation that live in the even (${\cal O}_+$) and the odd (${\cal
O}_-$) sectors, respectively.

\subsection{The ``new Dirac equation''}

The ``new Dirac equation'' (NDE) proposed by Dirac \cite{Dirac}
four decades after Majorana's work, reads, in our conventions
\footnote{The correspondence of Dirac's notations
\cite{Dirac} with ours is $q_a=L_a$,
$\alpha^0=(\gamma^0)^{ab}, \alpha^1=(\gamma^2)^{ab}, \alpha^2=(\gamma^3)^{ab}$
and $\alpha^3=(\gamma^1)^{ab}$.}
,
\begin{equation}\label{Dirac}
D_a|\Psi (x)\rangle=0,
\quad\hbox{where}\quad
 D_a=(-iP^\mu\gamma_\mu+m)_a\,^bL_b.
\end{equation}
The formal similarity with the usual spin one-half Dirac equation
is merely superficial~: $|\Psi(x)\rangle$ here   is an infinite
component field (due to its expansion in  Fock space), and has no
spinor index $a$.

Contracting $D_a$ operator in (\ref{Dirac}) with
$L^b(\lambda)_b\,^a$, where $(\lambda)_a\,^b$ is an arbitrary
$4\times4$ matrix, we obtain fifteen independent consistency
equations (for $\lambda=\gamma_0\gamma_1\gamma_2\gamma_3$
contraction gives the identity $0=0$) that can be organized as
follows.
\begin{eqnarray}
&(P^\mu\Gamma_\mu-\frac{1}{2}m)|\Psi(x)\rangle=0\, ,&\label{maj1}
\\[6pt]
&(m\Gamma_\mu+\frac{1}{2}P_\mu-iS_{\mu\nu}P^\nu)|\Psi(x)\rangle=0\,
,&\label{Sta0}
\\[8pt]
&(\Gamma_{\mu}P_\nu-\Gamma_{\nu}P_\mu+imS_{\mu\nu})|\Psi(x)\rangle=0\,
,&
\\[8pt]
&W^\mu|\Psi (x)\rangle=0,\, &\label{W}
\end{eqnarray}
where $W^\mu$ is the Pauli-Lubanski vector. The Klein-Gordon
equation appears as a consistency condition, requiring the
commutator to vanish,
\begin{equation}
[D_a,D_b]|\Psi(x)\rangle=iC_{ab}(P^2+m^2)|\Psi(x)\rangle=0.
\label{DirKG}
\end{equation}

The Klein-Gordon equation (\ref{DirKG}) selects, out of  all
solutions of Majorana equation (\ref{maj1}), the one with the lowest
possible spin, $J=0$, as seen from the mass formula
(\ref{MassSpectr}) with $M=\frac{m}{2}$.
 The NDE (\ref{Dirac}), 
describes therefore a spinless massive particle of positive energy.
The solution of the Dirac equation in the internal space -
coordinate representation (i.e. $\psi_+(x,q)=\langle
q|\Psi_{J_+=0}(x)\rangle$) is
\begin{equation}\label{solD}
\psi_+(x,q)=A\exp\left\{\frac{-m(q_1^2+q_2^2)
-ip^2(q_1^2-q_2^2)+i2p^3q_1q_2}{2(p^0-p^1)}\right\} \exp{(ix^\mu
p_\mu)},
\end{equation}
where $A$ is an arbitrary constant. Since this is an even function
under internal reflection, we have
$$
{\cal R}\psi_+(x,q)=\psi_+(x,q).
$$
In the rest frame the solution reduces to the ground state of a
planar harmonic oscillator,
\begin{eqnarray}\label{solD:r}
\psi^{(0)}_+(t,q)&=&A\exp\left\{\frac{-(q_1^2+q_2^2)}{2}\right\}
\exp{(-itm)}
= A\langle q|00\rangle \exp{(-itm)}.
\end{eqnarray}

\subsection{The Staunton equation}

In 1974  Staunton \cite{Stau2} observed that  the Majorana and
Klein-Gordon equations can  both be obtained directly from the
consistency condition (\ref{Sta0}), instead of the original Dirac
equation, (\ref{Dirac}).
Then, Staunton's idea
 was to modify  (\ref{Sta0}) by putting an arbitrary coefficient,
$\kappa$,  in front of $P_{\mu}$. In his analysis, Staunton
arrived at the conclusion that the modified equation is consistent
with the Poincar\'e representation for only two values of this
parameter, namely for $\kappa=\frac{1}{2}$ and $\kappa=1$.
We express this as
\begin{eqnarray}\label{Stau2}
D^{(\kappa)}_{\mu} |\Psi(x)\rangle=0,\qquad
D^{(\kappa)}_{\mu}=m\Gamma_\mu+\kappa P_\mu-iS_{\mu\nu}P^\nu.
\end{eqnarray}
As consistency conditions, (\ref{Stau2}) implies the Klein-Gordon
and Majorana equations,
\begin{equation}\label{KGMaj}
(P^2+m^2)|\Psi(x)\rangle=0, \qquad\hbox{and}\qquad
 (P^\mu\Gamma_\mu-m\kappa)|\Psi(x)\rangle=0.
\end{equation}
We require that the commutator annihilates the physical states,
$$
[D_\mu,D_\nu]|\Psi(x)\rangle=\left[-i(P^2+m^2)S_{\mu\nu}-P_\mu
D_\nu +P_\nu D_\mu\right]|\Psi(x)\rangle=0.
$$
Then,  contracting this equation with $S^{\mu\nu}$ and  taking
into account the first relation from (\ref{C1C2V}), we find that
(\ref{Stau2}) implies the Klein-Gordon equation.

 The Majorana
equation appears in turn upon contracting (\ref{Stau2}) with
$P^\mu$ and using the Klein-Gordon equation.

From (\ref{KGMaj}) and (\ref{Wsq1}) we
get, for $\kappa=\frac{1}{2}$, $W^\mu W_\mu=0$. The spin is hence
zero, and (\ref{Stau2}) is
equivalent to the original Dirac equation (\ref{Dirac}). For
$\kappa=1$ we have, instead,
\begin{equation}
W^\mu W_\mu =m^2\,\frac{1}{2}\left(1+\frac{1}{2}\right),\\
\end{equation}
so that (\ref{Stau2}) describes a spin $\frac{1}{2}$ particle.

For $\kappa=1$, the general solution of (\ref{Stau2}) can be
expressed,
 in internal coordinate space,
 in terms of the solution (\ref{solD}) of the Dirac system,
\begin{equation}\label{solS}
\Psi_-(x,q)=(Bq_1+Cq_2)\Psi_+(x,q),
\end{equation}
where $B$, $C$ are arbitrary constants. Note that $\Psi_-(x,q)$ is
an odd function of $q_i$,
$$
{\cal R}\Psi_-(x,q)=-\Psi_-(x,q).
$$
In the rest frame, (\ref{solS}) reduces to the first exited state of
a planar harmonic oscillator,
\begin{eqnarray}\label{solS:r}
\Psi_-^{(0)}(t,q)&=&\left(Bq_1+C
q_2\right)\exp\left\{\frac{-(q_1^2+q_2^2)}{2}\right\}\exp{(-itm)}\\
&=&\left(B\langle q|10\rangle+C\langle
q|01\rangle\right)\exp{(-itm)}.\nonumber
\end{eqnarray}

In the Fock space, the rest frame solutions (\ref{solD:r}) and
(\ref{solS:r}) take the form
\begin{eqnarray}\label{restsol}
\begin{array}{lll}\label{sol+}
|\Psi_+^{(0)}(x)\rangle=\exp(-itm)\psi^{0}_{++}\vert 00
\rangle_{++},\qquad &\hbox{ spin}\quad 0,
\\[14pt]\label{sol-}
|\Psi_-^{(0)}(x)\rangle=\exp(-itm)\left( \psi^{0}_{+-}\vert 00
\rangle_{+-}+\psi^{0}_{-+}\vert 00 \rangle_{-+}\right),\qquad
&\hbox{ spin}\quad \frac{1}{2},
\end{array}
\end{eqnarray}
(see (\ref{O+}) for $J_+=0$, and (\ref{O-}) for $J_-=\frac{1}{2}$).
After an arbitrary Lorentz transformation, all states in the
corresponding Fock subspace ${\cal O}_+$ or ${\cal O}_-$ can be
occupied, cf. (\ref{psi+}) and (\ref{psi-}). All coefficients will
be
 linear combinations of the only independent coefficient
$\psi^{00}_{++}$ (spin $0$ case), or  of $\psi^{00}_{+-}$ and
$\psi^{00}_{-+}$ (spin $1/2$ case). Note here that
$$
S_z\vert 00 \rangle_{++}=0,\qquad S_z\vert 00
\rangle_{+-}=-\frac{1}{2}\vert 00 \rangle_{+-},\qquad S_z\vert 00
\rangle_{-+}=\frac{1}{2}\vert 00 \rangle_{-+}.
$$

\section{Our unified supersymmetric theory}\label{ourtheory}

We have seen that the Dirac and Staunton equations extract, via the
Klein-Gordon equation, the lowest spin states, namely $J=0$ and
$J=\frac{1}{2}$, respectively, from the Majorana spectrum
(\ref{maj}). Now we show how these two cases can be merged into a
single supersymmetric one.  We posit the equation
\begin{equation}
\left(D_\mu^{1/2}\Pi_++D_\mu^{1}\Pi_-\right)|\Psi(x)\rangle=0,
\label{ourequation1}
\end{equation}
where the $D_\mu^{\kappa}$'s are the operators in the Staunton
equation (\ref{Stau2}), and the $\Pi_\pm$ are the projectors
(\ref{projectors}). Then (\ref{ourequation1}) becomes
\begin{equation}
{\cal D}_\mu|\Psi(x)\rangle=0,\qquad {\cal
D}_\mu=m\Gamma_\mu+\hat{\kappa}P_\mu-iS_{\mu\nu}P^{\nu}.
\label{ourequation}
\end{equation}
Our equation (\ref{ourequation}) amounts hence
to promoting
Staunton's constant $\kappa$ to  an  \emph{operator},
$\hat{\kappa}$, on ${\cal O}$,
\begin{equation}\label{kappa}
   \hat{\kappa}=\frac{1}{4}(3-{\cal R}),
\end{equation}
whose eigenvalues are precisely those appropriate for the new
Dirac and Staunton equations,
\begin{equation}\label{kappabis}
   \hat{\kappa}\,|\Psi_+(x)\rangle=\frac{1}{2}|\Psi_+(x)\rangle, \qquad
   \hat{\kappa}\,|\Psi_-(x)\rangle=|\Psi_-(x)\rangle.
\end{equation}

Let us note, however, that the similarity of  (\ref{ourequation})
with Staunton's equation (\ref{Stau2}) is deceiving, in that the
operator $\hat\kappa$ is non-local in the internal
translation-invariant variables $q_i$, $i=1,2$. Moreover, since the
general solutions of (\ref{ourequation}) is an arbitrary combination
of (\ref{solD}) and (\ref{solS}), our equation activates
simultaneously the spin $0$ and spin $\frac{1}{2}$ fields.
Projecting $\Pi_+{\cal
D}_\mu|\Psi(x)\rangle=D_\mu^{1/2}|\Psi_+(x)\rangle$, we get the
spin-$0$ Dirac system reduced onto  ${\cal O}_+$ , and for
$\Pi_-{\cal D}_\mu|{\Psi}(x)\rangle=D_\mu^{1}|\Psi_-(x)\rangle$ we
get the spin-$\frac{1}{2}$ Staunton system reduced onto  ${\cal
O}_-$.
 In this way, our new equation describes a spin
$(0,\frac{1}{2})$ supermultiplet.
 Then consistency of our new
equation (\ref{ourequation}) implies, once again, a Klein-Gordon and
a Majorana  equations
\begin{equation}\label{ourKG}
(P^2+m^2)|\Psi(x)\rangle=0
\end{equation}
and
\begin{equation}\label{ourM}
(P^\mu\Gamma_\mu-m\hat{\kappa})|\Psi(x)\rangle=0,
\end{equation}
respectively. The first one fixes the mass, \textit{in both sectors}, to be
$m$. The mass term in the Majorana equation is now an operator
$M=m\hat{\kappa}$, which takes different values
in the even and odd subspaces of the Hilbert space,
\begin{eqnarray}
\begin{array}{llll}
\Pi_+(P^\mu\Gamma_\mu-m\hat{\kappa})|\Psi
(x)\rangle&=&(P^\mu\Gamma_\mu-\frac{1}{2}m)|\Psi_+(x)\rangle\,
,\qquad& \hbox{integer spin}\, ,
\\[10pt]
\Pi_-(P^\mu\Gamma_\mu-m\hat{\kappa})|\Psi
(x)\rangle&=&(P^\mu\Gamma_\mu-\;m)|\Psi_-(x)\rangle\, ,&
\hbox{half-integer spin}\, .
\end{array}
\label{masses}
\end{eqnarray}
The Klein-Gordon equation  (\ref{ourKG}) implies that the spin of
every solution in (\ref{masses}) is necessarily the lowest
possible one, namely zero for $|\Psi_+(x)\rangle$ and
$\frac{1}{2}$ for $|\Psi_-(x)\rangle$. This is  consistent with
the mass formula (\ref{MassSpectr}), yielding
the \emph{same} mass 
for the  fields $|\Psi_+(x)\rangle$ and $|\Psi_-(x)\rangle$,
$$
\frac{m/2}{0+1/2}=\frac{m}{1/2+1/2}=m
$$
cf. also (\ref{ourKG}).  Due to our specific representation
(\ref{so32g}) of the $SO(2,3)$ group, we have the relations
(\ref{id}), from which we obtain the identities
\begin{equation}\label{gaugearg}
W^\mu {\cal D}_\mu\equiv0, \qquad {\cal P}^\mu {\cal
D}_\mu\equiv0,
\end{equation}
where $W^\mu$ is the Pauli-Lubanski vector, and
\begin{equation}
{\cal P}_\mu=\frac{1}{2}P_\mu+(3\hat{\kappa}-1)(P\Gamma)\Gamma_\mu
+i\hat{\kappa}S_{\mu\nu}P^\nu.
\end{equation}
Relations (\ref{gaugearg}) indicate that only two components of
${\cal D}_\mu$ yield independent equations. Four components are
 necessary to assure the covariance of the equations.


Now we identify the supercharge operator. Let us consider the
Hermitian  $4$-component spinor  operator
\begin{equation}\label{Scharge}
Q_a=\frac{1}{\sqrt{m}}(-i{\cal R}P^\mu\gamma_\mu+m)_a\,^bL_b
\qquad a=1,\dots, 4,
\end{equation}
where the $L_a$ are those  internal
$osp(1|4)$ generators in (\ref{La}).
 This is an observable operator with respect our
equations,
\begin{equation}\label{Q,D_mu}
[{\cal D}_\mu,Q_a]=-\frac{i}{2m}{\cal
R}(P^2+m^2)(\gamma_\mu)_a{}^bL_b+(i\gamma_\mu-\frac{1}{m}P_\mu)_a{}^bD_b\Pi_+\approx0,
\end{equation}
and consequently, also with respect to the Klein-Gordon and the
Majorana equations. In (\ref{Q,D_mu})  $D_b$ is the Dirac operator
from (\ref{Dirac}); here and in what follows $\approx$ denotes
equality on the surface defined by the corresponding field equations.

This operator transforms the spin-0 particle into the
spin-$\frac{1}{2}$ particle, and vice versa. To show this, let
$|\Psi_\pm(x)\rangle=\Pi_\pm|\Psi (x)\rangle$ be solutions of
(\ref{ourequation}). Then, due to Eqs. (\ref{Pi,SGL}) and
(\ref{Q,D_mu}), we have $Q_a\Pi_\pm=\Pi_\mp Q_a,$ and
$$Q_a|\Psi_\mp(x)\rangle \approx |\Psi_\pm(x)\rangle  .$$
 It is
illustrative to verify this in the rest frame, where
\begin{eqnarray}\label{Q0r}
Q^{(0)}_a|\Psi_\pm^{(0)}(x)\rangle=\sqrt{2m}\,(a^\pm_1,a^\pm_2,\pm
ia^\pm_1,\pm ia^\pm_2)|\Psi_\pm^{(0)}(x)\rangle.
\end{eqnarray}
With (\ref{restsol}), Eq. (\ref{Q0r}) yields
$$
Q^{(0)}_a|\Psi_+^{(0)}(x)\rangle\approx|\Psi_-^{(0)}(x)\rangle,
\qquad
Q^{(0)}_a|\Psi_-^{(0)}(x)\rangle\approx|\Psi_+^{(0)}(x)\rangle.
$$

The $Q_a$ operator satisfies non-linear anticommutation relations,
\begin{eqnarray}
\{Q_a,Q_b\}=&&
(-3P^\mu+Z^\mu)(\gamma_{\mu})_{ab}%
-4imZ^{\mu\nu}(\gamma_{\mu\nu})_{ab} \label{QQ*}
\\[12pt]
&&+\frac{2}{m}(P^2+m^2) (iS_{\mu\nu}\gamma^{\mu\nu}+4i\frac{1}{P^2}
P_\mu S_{\nu\lambda}P^\lambda\gamma^{\mu\nu}+\Gamma_\mu\gamma^{\mu})_{ab}%
-\frac{4}{m}(P\Gamma-m\hat{\kappa})(\gamma_{\mu} P^\mu)_{ab},
\nonumber
\end{eqnarray}
where
\begin{equation}\label{Z}
Z^\mu=-{\cal R}P^\mu,\qquad Z^{\mu\nu}=\pi^\mu{}^\rho
\pi^\nu{}^\lambda S_{\rho \lambda},
\end{equation}
and
\begin{equation}
\pi_{\mu\nu}=\eta_{\mu\nu}-\frac{P_\mu P_\nu}{P^2}\approx
\eta_{\mu\nu}+\frac{P_\mu P_\nu}{m^2}.
\end{equation}
We note that $Z_{\mu\nu}$ is a covariant expression of the spin
operator $S_{ij}$. (In the rest frame, the projectors reduce to
$\pi_{\mu\nu}=(0,\delta_{ij})$ so that $Z_{\mu\nu}$ reduces to
$S_{ij}$.)  Note also that on shell, the first term in the
anticommutator (\ref{QQ*}),
 $(-3P^\mu+Z^\mu)\gamma_\mu=-(3+{\cal R})(P\gamma)$,
 is positive  definite.

The terms in the second line in (\ref{QQ*})  include, as commuting
factors, the Klein-Gordon and the Majorana operators. Putting them
to zero, the on-shell anticommutator is obtained,
\begin{eqnarray}\label{QQ2}
\{Q_a,Q_b\}&\approx&\left(-3P^\mu+Z^\mu\right)(\gamma_{\mu})_{ab}-4im
Z^{\mu\nu}(\gamma_{\mu\nu})_{ab}.
\end{eqnarray}
 The translation and Lorentz generators $P_\mu$ and ${\cal
J}_{\mu\nu}$, together with the supercharge $Q_a$, obey the
commutation relations
\begin{eqnarray}\label{salg}
&[{\cal J}_{\mu\nu},{\cal J}_{\lambda\rho}]=i(\eta_{\mu\lambda}{\cal
J}_{\nu\rho} + \eta_{\nu\rho}{\cal J}_{\mu\lambda}-
\eta_{\mu\rho}{\cal J}_{\nu\lambda}- \eta_{\nu\lambda}{\cal
J}_{\mu\rho}),&
\\[12pt]
&[{\cal J}_{\mu\nu},P_\lambda]=i(\eta_{\mu\lambda} P_\nu-
\eta_{\nu\lambda}P_\mu), \qquad [P_\mu,P_\nu]=0,&\nonumber
\\[12pt]
&[{\cal J}_{\mu\nu},Q_a]=-(\gamma_{\mu\nu})_a\,^bQ_b, \qquad
[P_\mu,Q_a]=0.&\label{JQ}
\end{eqnarray}
The  operators $P_\mu$, ${\cal J}_{\mu\nu}$ and $Q_a$, together with
the operators $Z_\mu$ and $Z_{\mu\nu}$ appearing in the
anticommutator of the supercharge, form the set of symmetry
generators for our system,
\begin{equation}\label{A}
[{\cal D}_\mu, {\cal A}]\approx0,\qquad \hbox{for}\quad {\cal
A}={\cal J}_{\mu\nu},\,{}P_\mu,\,{}Q_a,\,{}Z_\mu,\,{}Z_{\mu\nu}.
\end{equation}
The reflection operator plays, in our superalgebraic structure, the
r\^ole of the grading operator,
\begin{equation}
[{\cal R},{\cal J}_{\mu\nu}]=0, \qquad [{\cal R},P_\mu]=0, \qquad
[{\cal R},Z_\mu]=0,\qquad [{\cal R},Z_{\mu\nu}]=0,\qquad \{{\cal
R},Q_a\}=0. \label{Ralg}
\end{equation}
However here, unlike the 2+1-dimensions \cite{anyonsusy},  we do
not have a Lie superalgebraic structure on-shell, since, although
$Z_\mu$ and $Z_{\mu\nu}$ are translationally invariant vector
resp. antisymmetric tensor operators, their commutators with  the
supercharge  $Q_a$ are nontrivial, nonlinear in symmetry
generators,
\begin{equation}\label{ZZQ}
    [Z_\mu,Q_a]=2Z_\mu Q_a,\qquad
    [Z_{\mu\nu},Q_a]=-\pi_{\mu\lambda}\pi_{\nu\rho}
    (\gamma^{\lambda\rho})_a{}^bQ_b.
\end{equation}
  Note also that here
\begin{eqnarray}\label{ZZ}
\begin{array}{c}
[Z_{\mu},Z_{\nu}]=0,\quad[Z_{\mu},Z_{\nu\lambda}]=0, \\[12pt]
[Z_{\mu\nu},Z_{\lambda\rho}]=i(\pi_{\mu\lambda}Z_{\nu\rho}
+\pi_{\nu\rho}Z_{\mu\lambda}-
\pi_{\mu\rho}Z_{\nu\lambda}-\pi_{\nu\lambda}Z_{\mu\rho}),
\end{array}
\end{eqnarray}
to be compared with the tensorial extensions which appear in
supergravity, and for superbranes
\cite{Bergshoeff:1987cm}-\cite{Gauntlett:2000ch}.

In spite of these complications due to nonlinearity, the invariant
operator playing the r\^ole of the Casimir operator is easily
identified~: up to the $m^2$ factor, it can be namely identified
as the superspin (see below),
\begin{eqnarray}\label{superspin}
{\cal C}=W^\mu W_\mu -\frac{1}{64}\chi^\mu \chi_\mu,\qquad [{\cal
C}, {\cal A}]\approx0,
\end{eqnarray}
where $\chi_\mu=Q^a(\gamma_\mu)_a{}^bQ_b$. On-shell it takes the
value
$$
{\cal C}=m^2.
$$

We also note that although we have a non-linear,
W-type \cite{Boer},
symmetry super-algebra, the Jacobi identities are valid owing to the
associativity of  all involved operator products.

\section{Supersymmetric higher-spin Majorana-Klein-Gordon system}
\label{Majoranasusy}

As we already noted, the main properties of the Majorana equation
strongly  suggest that some kind of supersymmetry could
be  involved.  We have also shown that,
for the lowest (namely the zero and one-half)
 spin states, a supersymmetric theory
can indeed be constructed. It is therefore natural to ask if it is
possible to extend the supersymmetry to some arbitrary-spin massive
supermultiplet. A priori, we know that $P^2$ should be a
Casimir operator. Requiring
supersymmetry, we expect the appearance of the Klein-Gordon
equation as a consistency condition. In fact, when we impose,
simultaneously,
\begin{equation}\label{KGM:J}
  (P^2+m^2)|\Psi_J(x)\rangle=0,
\qquad
 (P^\mu\Gamma_\mu-M_J)|\Psi_J(x)\rangle=0
\end{equation}
where $M_J=(J+\frac{1}{2})m$, $J=0,\frac{1}{2},1,\frac{3}{2},...$,
we extract from the infinite spectrum of the Majorana equation an
irreducible representation of the  Poincar\'e group with the mass
$m$, spin $J$ and positive energy.

Now, we extend the supersymmetry of the spin $(0,\frac{1}{2})$
supermultiplet constructed above to an arbitrary spin $(J_+,J_-)$
supermultiplet  such that
 $|J_+-J_-|=\frac{1}{2}$.  By convention, $J_+$ is integer  and $J_-$ is half integer .

Firstly, we generalize the Majorana equation (\ref{maj}) to
\begin{equation}\label{Maj:gen}
(P^\mu\Gamma_\mu-\hat{M}_{\cal S})|\Psi(x)\rangle=0,
\end{equation}
where  the mass parameter has been traded for an operator
cf. (\ref{ourM}),
\begin{eqnarray}\label{M_S}
&\hat{M}_{\cal S}=\frac{m}{2}\left(J_++J_-+1+(J_+-J_-){\cal
R}\right),&\\
&J_+=0, 1, \ldots,\quad J_-=\frac{1}{2}, \frac{3}{2}, \ldots,\qquad
|J_+-J_-|=\frac{1}{2}\, .&\label{JJJ}
\end{eqnarray}
Projected to the even and odd Fock subspaces, this equation is
equivalent to
\begin{equation}
\Pi_\pm(P^\mu\Gamma_\mu-\hat{M}_{\cal S})|\Psi(x)\rangle=
(P^\mu\Gamma_\mu-M_{J_\pm})|\Psi_\pm(x)\rangle=0,
\end{equation}
where $|\Psi_\pm(x)\rangle=\Pi_\pm|\Psi (x)\rangle$, see
(\ref{R:psi}). By construction, the solution of (\ref{Maj:gen}) is
the sum of $|\Psi_+(x)\rangle$ and $|\Psi_-(x)\rangle$ belonging
 to the integer (resp. half-integer) spin subseries
of solutions of the Majorana equation (\ref{maj}).
 It follows from the mass formula (\ref{MassSpectr}) that the
states with spins $J_+$ and $J_-$ have again equal masses, namely
$m$. (\ref{Maj:gen}) has a  supermultiplet in its spectrum
therefore. The  equations which describe it read
\begin{equation}\label{KGM:SUSY}
  (P^2+m^2)|\Psi_{\cal S}(x)\rangle=0,
\qquad
 (P^\mu\Gamma_\mu-\hat{M}_{\cal S})|\Psi_{\cal S}(x)\rangle=0.
\end{equation}
Their spin content is
\begin{equation}\label{spincontent}
\left({\cal S}-\frac{\triangle J}{2},{\cal S}+\frac{\triangle
J}{2}\right),\qquad {\cal S}=\frac{J_++J_-}{2},\qquad \triangle
J=J_--J_+=\pm \frac{1}{2}\, .
\end{equation}

The observable hermitian supercharge is
\begin{equation}\label{Scharge+-}
Q^{(\pm)}_a=\frac{1}{\sqrt{m}}(\mp i{\cal
R}P^\mu\gamma_\mu+m)_a\,^bL_b, \qquad \triangle J=\pm\frac{1}{2}.
\end{equation}
Its commutator with the equations (\ref{KGM:SUSY}) vanishes
on-shell,
\begin{equation}\label{KGM,Q:comm.}
[P^\mu\Gamma_\mu-\hat{M}_{\cal S},Q^{(\pm)}_a]=\pm\frac{{\cal
R}L_a}{2}(P^2+m^2)\approx 0,\qquad [(P^2+m^2),Q^{(\pm)}_a]= 0.
\end{equation}
The first equation from  (\ref{KGM,Q:comm.}) means that
supersymmetry itself requires satisfying the Klein-Gordon
equation. Hence,  the $Q_a$ is an (observable) odd symmetry
generator for the  equations (\ref{KGM:SUSY}). Being a spinor
operator, it intertwines physical states of spin $J_+$ and of spin
$J_-$.

On-shell (given by equations (\ref{KGM:SUSY})), instead of
(\ref{QQ2}), we obtain the anticommutation relation
\begin{eqnarray}\label{QQ:superS}
&\{Q^{(\pm)}_a,Q^{(\pm)}_b\}\approx-2(1+2{\cal
S})P^\mu(\gamma_{\mu})_{ab}\pm
Z^\mu(\gamma_{\mu})_{ab}-4imZ^{\mu\nu}(\gamma_{\mu\nu})_{ab}, \qquad
\triangle J=\pm \frac{1}{2},&
\end{eqnarray}
where
$$
{\cal S}=\frac{1}{4},\frac{5}{4},\frac{9}{4},\dots, \quad \hbox{for}
\quad\triangle J= \frac{1}{2},\quad \hbox{and}\quad {\cal
S}=\frac{3}{4},\frac{7}{4},\frac{11}{4},\dots, \quad \hbox{for}
\quad\triangle J=- \frac{1}{2},
$$
and  $Z's$ are the same as in (\ref{Z}).

The form of the superalgebraic structure (\ref{QQ2})--(\ref{JQ}),
(\ref{Ralg})--(\ref{ZZ}), with (\ref{QQ2}) exchanged for
(\ref{QQ:superS}), is preserved. The $Z's$ here  are the conserved
charges with respect to equations (\ref{KGM:SUSY}). The invariant
operator related to superspin is now
\begin{eqnarray}
{\cal C}^+=W^\mu W_\mu -\frac{1}{64}\chi^\mu \chi_\mu,\qquad \triangle J=+ \frac{1}{2},\\[12pt]
{\cal C}^-=W^\mu W_\mu +\frac{1}{3\cdot64}\chi^\mu \chi_\mu, \qquad
\triangle J=- \frac{1}{2}.
\end{eqnarray}
It takes the on-shell values
\begin{eqnarray}
{\cal C}^+=2m^2\left({\cal S}+\frac{3}{4}\right)\left({\cal
S}+\frac{1}{4}\right),\qquad {\cal S}=\frac{1}{4}, \frac{5}{4},\frac{9}{4}, \dots,\\[12pt]%
{\cal C}^-=\frac{2m^2}{3}\left({\cal
S}-\frac{1}{4}\right)\left({\cal S}+\frac{5}{4}\right),\qquad
{\cal S}=\frac{3}{4}, \frac{7}{4}, \frac{11}{4},\dots\,.
\end{eqnarray} In this way, the Majorana-Klein-Gordon
(\ref{KGM:SUSY}) system describes, universally, a massive
super-multiplet of spin content $(J_+,J_-)=\left({\cal
S}-\frac{\triangle J}{2},{\cal S}+\frac{\triangle J}{2}\right)$.
Our
previous results are plainly recovered for ${\cal S}=\frac{1}{4}$ and
$\triangle J=\frac{1}{2}$.

\subsection{The large superspin limit}

It is interesting to study the behavior of our superalgebra for
large values of the super-spin. Let us first redefine the
supercharges (\ref{Scharge+-}),
\begin{equation}
{\cal Q}^{(\pm)}_a=\frac{1}{\sqrt{1+2{\cal S}}}Q^{(\pm)}_a.
\end{equation}
Off-shell they satisfy the anticommutation relation
\begin{eqnarray}\label{calQQ}
&&\{{\cal Q}^{(\pm)}_a,{\cal
Q}^{(\pm)}_b\}=-2P^\mu(\gamma_{\mu})_{ab} \pm\frac{1}{1+2{\cal
S}}Z^\mu(\gamma_{\mu})_{ab} -\frac{4im}{1+2{\cal
S}}Z^{\mu\nu}(\gamma_{\mu\nu})_{ab} \label{QQ}
\\[12pt]
&&+\frac{2(P^2+m^2)}{m(1+2{\cal S})}
\left(iS_{\mu\nu}\gamma^{\mu\nu}+4i\frac{1}{P^2} P_\mu
S_{\nu\lambda}P^\lambda\gamma^{\mu\nu}+\Gamma_\mu\gamma^{\mu}\right)_{ab}
-\frac{4}{m(1+2{\cal S})}(P\Gamma-\hat{M}_{\cal S})(\gamma_{\mu}
P^\mu)_{ab}, \nonumber
\end{eqnarray}
while the commutator of $Z_\mu$ and $Z_{\mu\nu}$ with ${\cal
Q}^{(\pm)}_a$ remains of the form (\ref{ZZQ}). When ${\cal
S}\rightarrow\infty$, (\ref{calQQ})  takes the usual form of
the $N=1$
supersymmetric anticommutation relation,
\begin{eqnarray}\label{QQ:SPoincare}
&\{{\cal Q}^{(\pm)}_a,{\cal
Q}^{(\pm)}_b\}=-2P^\mu(\gamma_{\mu})_{ab}.
\end{eqnarray}
On the other hand, defining
$$
\qquad W_{\mu\nu}=-\frac{4m}{1+2{\cal S}}Z_{\mu\nu}
$$
we get, in this limit,
\begin{eqnarray}\label{QQ:SPoinc.C-Ext}
&\{{\cal Q}^{(\pm)}_a,{\cal
Q}^{(\pm)}_b\}=-2P^\mu(\gamma_{\mu})_{ab}+W^{\mu\nu}(\gamma_{\mu\nu})_{ab},
\quad\hbox{and}\quad[W_{\mu\nu},{\cal Q}^{(\pm)}_a]=0.
\end{eqnarray}
 Using (\ref{ZZ}), we have
$$
[W_{\mu\nu},W_{\lambda\rho}]=\frac{4m}{1+2{\cal
S}}(\pi_{\mu\lambda}W_{\nu\rho} +\pi_{\nu\rho}W_{\mu\lambda}-
\pi_{\mu\rho}W_{\nu\lambda}-\pi_{\nu\lambda}W_{\mu\rho}),
$$
and in the limit ${\cal S}\rightarrow\infty$ we find that the
$W_{\mu\nu}$ turns into an abelian tensorial central charge.

So, with (\ref{QQ:SPoincare}) or (\ref{QQ:SPoinc.C-Ext}), off-shell
we obtain the super-Poincar\'e algebra without or with tensorial
central extension. Note that in the large-spin limit the
nonlinearity in the superalgebraic structure disappears off-shell.

Remember that our construction includes, from the beginning,
 a hidden  length parameter $l$, used to transform the
canonical operators $q_i$ and $\eta_i$ into dimensionless 2D
Heisenberg generators.   This parameter
can be identified with the AdS radius. Then we note also here that
the (un)extended Poincar\'e superalgebra we have gotten can be
obtained as a limit of the $osp(1|4)$, when the AdS radius tends to
infinity \cite{Bandos:1999pq,Bandos:1999rp}.

\section{Concluding remarks and outlook}\label{conc}

We have constructed the covariant (3+1)D vector set of linear
differential equations, which describe  supermultiplets of spins 0
and 1/2 fields. In this  theory, the spin degrees of freedom are
carried by an internal 2D Heisenberg algebra. Extending the
construction to get a supermultiplet of spins $(j,j+1/2)$,
requires, however, to use a modified, first order  Majorana
equation, augmented with the second order Klein-Gordon equation.

Our results here can be compared with those in the (2+1)D case
\cite{Plyushchay:1994re,CortesPlyushchay,Plyushchay:1993ak,Plyu:Spinorial
Eq.}, where an analogous supersymmetric  construction has been
carried out in \cite{anyonsusy}. It is based on the 1D deformed
Heisenberg algebra with reflection \cite{Plyushchay:1997ty}, $
\big[a^-,a^+\big]=1+\nu R,$ $ \{a^\pm,R\}=0,$ $R^2=1,$ and involves
the $osp(1|2)$ superalgebra that allowed us to describe,
universally, either an anyonic supermultiplet of spins $\pm
(s,s+1/2)$ or a supermultiplet, $(j,j+1/2)$, of usual fields of
integer and half-integer spin. In the former case,
$s=\frac{1}{4}(1+\nu)>0$ can take arbitrary real values for the
unitary infinite dimensional representations of the deformed
Heisenberg algebra, characterized by the deformation parameter
values $\nu>-1$. The latter case arises when we choose
finite-dimensional non-unitary representations of the algebra
corresponding to the negative odd values of $\nu$
\cite{Plyushchay:1997ty}. The underformed Heisenberg algebra
($\nu=0$) gives a semionic supermultiplet $(1/4,3/4)$
\cite{Sorokin:1992sy}.

In (2+1) dimensions spin is a pseudoscalar, and both members of the
supermultiplet have, on-shell, the same number of spin degrees of
freedom (namely equal to one). As a consequence, there, on shell,
appears a usual Poincar\'e superalgebra. In the present (3+1)D case,
the integer and half integer spin members of a supermultiplet are
described on-shell by different numbers of spin components (cf.
\cite{Volkov:1973ix,Forste:1996aj}), and on-shell we have a
nonlinear superalgebra, that only in the large superspin limit
reduces to the Poincar\'e superalgebra with or without tensorial
central charge.

By an appropriate generalization of the Majorana-Klein-Gordon theory
presented in Section 5, one can obtain a bosonized supersymmetric
system with a more general, exotic  supermultiplet that includes
fields of  spins shifted by $n+\frac{1}{2}$, $n=1,2,\ldots$. In such
a generalized theory, unlike in the case $n=0$ considered here, the
supercharge will be a covariant object of spin $n+\frac{1}{2}$ of
the order $2n+1$ in the space-time translation generator $P_\mu$,
and will generate  some more complicated superalgebra.  Such a
generalization will be considered elsewhere.

Dirac observed that his new equation is inconsistent with the usual
minimal U(1) gauge coupling \cite{Dirac,BiedHan,Sudarshan:1981hf},
namely, that consistency requires  $F_{\mu\nu}=\partial_\mu
A_\nu-\partial_\nu A_\mu=0$. Hence, the  electromagnetic field can
only be a pure gauge. The aim of Staunton has been precisely to find
an improved theory which would remove this inconsistency~: coupling
the particle to a gauge field is essential. However, while he shows
that his theory does not produce immediately the obstruction
$F_{\mu\nu}=0$, the consistency, in fact, was not proved  in
\cite{Stau2}
\footnote{The commutator of the interacting Majorana and
Klein-Gordon equations (58.b) and (59) from \cite{Stau2} produces a
new, missing, nontrivial condition that includes a derivative term
$\partial_\lambda F_{\mu\nu}$, and the checking process should
continue.}.
This -- fundamental -- question 
still remains open.

Staunton's ideas have also been extended to a curved background
\cite{Ahner}. The generalization of our theory presented here to
interactions with gauge fields and gravity deserves separate study.

Our theory has a close relationship with the ``supertwistor''
approach that is used, in particular,  for the description of
higher-spin massless fields \cite{Bandos:1999pq,Bandos:1999rp},
based on the fundamental representation of $osp(1|4)$ (see also
\cite{Fradkin:1987ah} for a related approach). Indeed, our Majorana
spinor $L_a$ generates, like a twistor $\lambda_\alpha$ does, the 2D
Heisenberg algebra (\ref{Cab}), $[L_a,L_b]=iC_{ab}$, and satisfies
the relation $L^aL_a=2i$ of the form of the helicity constraint in
the twistor theory. Moreover, our vector equation
(\ref{ourequation}) has a form (see Eq. (\ref{twisteqs}) below)
similar to the twistor relation $P_\mu=\lambda\gamma_\mu\lambda$,
generating the mass zero constraint. In the supertwistor approach,
in the simplest case, the Grassmann-even twistor variable
$\lambda_\alpha$ is combined with the scalar Clifford algebra
generator $\psi$, $\psi^2=1$, to realize the $osp(1|4)$-odd
generator as a product $Q_\alpha=\psi\lambda_\alpha$. There the
r\^ole of the grading operator is played by the external,
Grassmann-even operator, anticommuting with $\psi$. In our case
instead, the reflection operator ${\cal R}$ is identified as the
grading operator
\footnote{Our theory is different from the classical, related
approach to higher-spin massless fields \cite{Frond} based on
higher-rank symmetric Lorentz tensors, see
\cite{ouvry,Vasiliev:1995dn} and references therein. Here, the
supersymmetric higher spin fields are massive, and the spin
degrees of freedom are hidden in the internal Fock space (cf. also
\cite{Buch}).}.

It is also instructive to compare the equations of Majorana, Dirac
and Staunton rewritten in the form
\begin{equation}\label{twisteqs}
\begin{array}{cc}
\frac{1}{4}P^\mu(\gamma_\mu)^{ab}L_aL_b+M=0, & \hbox{Majorana,} \\[12pt]
mL^a-iP^\mu(\gamma_\mu)^{ab}L_b=0, & \hbox{NDE,}\\[12pt]
\kappa P_\mu-\frac{m}{4}(\gamma_\mu)^{ab}L_a L_b-\frac{1}{2}P^\nu
(\gamma_{\mu \nu})^{ab}L_a L_b=0, & \hbox{Staunton,}
\end{array}
\end{equation}
with the constraints appearing in the twistor formulation of
massive spin fields, see e.g.
\cite{Fedoruk:2003td}-\cite{Bars:2007vs}.

So, it would be interesting to work out in more details the relation
of our bosonized supersymmetry with the usual one in the
supertwistor approach.

Let us note here that a kind of ``a generalization of
global supersymmetry" \cite{Biedenharn1979} and a ``bosonic
counterpart of supersymmetry" \cite{Fedoruk:2005sg} were discussed
earlier in the literature in the context of the massive spin
theory. The approaches of van Dam-Biedenharn, of Fedoruk-Lukierski
and our present one here share the common feature that all three
theories are constructed in terms of infinite dimensional
representation of the Lorentz group, and involve internal, bosonic
twistor-like variables [cf. also \cite{Plyushchay:1992} for the
massless case]. Unlike our case, the models
\cite{Biedenharn1979,Fedoruk:2005sg} are characterized by an
infinite number of physical states of integer and half-integer
spin $J=0,\frac{1}{2},\ldots,$ which lie either on a linear Regge
trajectory $m_J^2\propto J+\frac{1}{2}$ \cite{Biedenharn1979}, or
have a fixed mass $m^2=const$ \cite{Fedoruk:2005sg}, cf. the
Majorana spectrum (\ref{MassSpectr}) $m_J^2\propto
(J+\frac{1}{2})^{-2}$. The essential difference is, however, that
in the approaches \cite{Biedenharn1979,Fedoruk:2005sg} the
Poincar\'e algebra is extended by a spinorial \emph{even}
operator. The latter interchanges integer and half-integer spin physical
states and satisfies \emph{commutation} relations [cf.
\cite{Plyushchay:1992}. \emph{Additional} Lorentz-scalar,
topologically nontrivial isospin variables transmute, after
quantization, the even spinorial integrals of motion into odd
supercharges]. A remarkable property of the theory in
\cite{Biedenharn1979} is that its spinorial charge involves a
space-time nonlocal operator that changes not only the spin, but
also the mass of the physical states consistently with the
Regge character of the spectrum. In \cite{Fedoruk:2005sg}, $P^2$,
instead, plays a role of the Casimir of the extended Poincar\'e
symmetry \emph{algebra}. In our case the bosonized supersymmetry
is characterized by a nonlinear \emph{superalgebra}, realized on
a finite supermultiplet, extracted from the infinite Majorana
spectrum.


In summary, we have shown that the Majorana-Dirac-Staunton theory
possesses a rich structure that allowed us to construct its
supersymmetric generalization without introducing any (Grassmann
odd, fermionic) additional spin degrees of freedom. Our
supersymmetric generalization relies on  non-locality in the
internal, translation invariant (twistor-like) bosonic variables.
The superalgebraic structure we obtain admits a nontrivial internal
symmetry, namely $Z_{\mu\nu}$. This nonlocality is similar to that
in other  bosonization constructions, where fermions are described
in terms of bosonic variables \cite{Naka}. Such  kind of
boson-fermion relation  is, in turn,  rooted in the underlying  nontrivial
topology, see e.g. \cite{Boson}--\cite{Houart}. We hope that
investigation of the field systems like those presented here could
reveal further connections between supersymmetry and topology.

\vskip 0.4cm\noindent {\bf Acknowledgements}. The authors thank V.
Akulov and D. Sorokin for valuable communications. MV is indebted to
the Laboratoire de Math\'ematiques et de Physique Th\'eorique of
Tours University for a hospitality extended to him. The work was
supported in part by FONDECYT (project 1050001) and by MECESUP
USA0108.

\appendix

\section*{Appendix}

 The generators (\ref{so32g}) are, explicitly,
\begin{eqnarray}\nonumber
S^{01}=-\frac{i}{4}(a_1^+\,^2+a_2^+\,^2-a_1^-\,^2-a_2^-\,^2),\qquad
S^{02}=-\frac{1}{4}(a_1^+\,^2-a_2^+\,^2+a_1^-\,^2-a_2^-\,^2),
\\[6pt]
S^{03}=\frac{1}{2}(a_1^+a_2^++a_1^-a_2^-),\qquad
S^{12}=\frac{1}{2}(N_1-N_2),\nonumber
\\[6pt]
S^{13}=-\frac{1}{2}(a_1^+a_2^-+a_1^-a_2^+),\qquad
S^{23}=\frac{i}{2}(a_1^+a_2^--a_1^-a_2^+),\nonumber
\\[6pt]
\Gamma^{0}=-\frac{1}{2}(N_1+N_2+1),\qquad
\Gamma^{1}=\frac{1}{4}(a_1^+\,^2+a_2^+\,^2+a_1^-\,^2+a_2^-\,^2),\nonumber
\\[6pt]
\Gamma^{2}=-\frac{i}{4}(a_1^+\,^2-a_2^+\,^2-a_1^-\,^2+a_2^-\,^2),\qquad
\Gamma^{3}=\frac{i}{2}(a_1^+a_2^+-a_1^-a_2^-).\nonumber
\end{eqnarray}

We can check that
\begin{eqnarray}\nonumber
   &S^{\mu\nu}S_{\mu\nu}=-\frac{3}{2}\, , \qquad
   \Gamma^\mu\Gamma_\mu=\frac{1}{2}\, ,&
   \\[8pt]
   &S_iS_i=\frac{1}{2}S_{ij}S_{ij}=\frac{N_1+N_2}{2}\left(\frac{N_1+N_2}{2}+1\right)
   .&\nonumber
\end{eqnarray}


\end{document}